\documentclass[letterpaper,aps,prd,twocolumn,
tightenlines,preprintnumbers,nofootinbib,showkeys,superscriptaddress]{revtex4-1}
\pdfoutput=1

\usepackage{fullpage}
\usepackage{amsfonts}
\usepackage{amsmath}
\usepackage{slashed}
\usepackage{amssymb}
\usepackage{graphicx}
\usepackage{makeidx}
\usepackage{cancel}
\usepackage{epic}
\usepackage{eepic}
\usepackage{epsfig}
\usepackage{latexsym}
\usepackage[usenames,dvipsnames]{xcolor}
\usepackage{float}
\usepackage{multirow, makecell}
\usepackage[export]{adjustbox}
\usepackage{xurl,hyperref}
\usepackage{enumitem}
\hypersetup{colorlinks=true,citecolor=red,linkcolor=NavyBlue,urlcolor=NavyBlue}
\usepackage[utf8]{inputenc}
\usepackage[caption=false]{subfig}
 
\usepackage{natbib}
\usepackage{relsize}
\usepackage[left=1.6cm,right=1.6cm,top=1.6cm,bottom=1.8cm]{geometry}
\usepackage{mathptmx}

\begin{document}
\relscale{1.05}
\captionsetup[subfigure]{labelformat=empty}

\title{Leptoquark-assisted Singlet-mediated Di-Higgs Production at the LHC}

\author{Arvind Bhaskar}
\email{arvind.bhaskar@research.iiit.ac.in}
\affiliation{Center for Computational Natural Sciences and Bioinformatics, International Institute of Information Technology, Hyderabad 500 032, India}

\author{Debottam Das}
\email{debottam@iopb.res.in}
\affiliation{Institute of Physics, Sachivalaya Marg, Bhubaneswar 751 005, India}
\affiliation{Homi Bhabha National Institute, Training School Complex, Anushakti Nagar, Mumbai 400 085, India}

\author{Bibhabasu De}
\email{bibhabasu.d@iopb.res.in}
\affiliation{Institute of Physics, Sachivalaya Marg, Bhubaneswar 751 005, India}
\affiliation{Homi Bhabha National Institute, Training School Complex, Anushakti Nagar, Mumbai 400 085, India}

\author{Subhadip Mitra}
\email{subhadip.mitra@iiit.ac.in}
\affiliation{Center for Computational Natural Sciences and Bioinformatics, International Institute of Information Technology, Hyderabad 500 032, India}

\author{Aruna Kumar Nayak}
\email{nayak@iopb.res.in}
\affiliation{Institute of Physics, Sachivalaya Marg, Bhubaneswar 751 005, India}
\affiliation{Homi Bhabha National Institute, Training School Complex, Anushakti Nagar, Mumbai 400 085, India}

\author{Cyrin Neeraj}
\email{cyrin.neeraj@research.iiit.ac.in}
\affiliation{Center for Computational Natural Sciences and Bioinformatics, International Institute of Information Technology, Hyderabad 500 032, India}

\date{\today}


\begin{abstract}
\noindent
At the LHC, the gluon-initiated processes are considered to be the primary source of di-Higgs production. However, in the presence of a new resonance, the light-quark initiated processes can also contribute significantly. In this letter, we look at the di-Higgs production mediated by a new singlet scalar. The singlet is produced in both quark-antiquark and gluon fusion processes through loops involving a scalar leptoquark and right-handed neutrinos.  With benchmark parameters inspired from the recent resonant di-Higgs searches by the ATLAS collaboration, we examine the prospects of such a  resonance in the TeV-range at the High-Luminosity LHC (HL-LHC) in the $b\bar{b} \tau^{+}\tau^{-}$ mode with a multivariate analysis. We obtain the $5\sigma$ and $2\sigma$ contours and find that a significant part of the parameter space is within the reach of the HL-LHC.
\end{abstract}
	
\maketitle

\section{Introduction}
\noindent
The current measurements of the Higgs boson cross section and its couplings agree with the Standard Model (SM) predictions. Still, new physics (NP) could be hiding in the Higgs sector if it primarily affects the Higgs couplings with the third generation quarks and vector bosons up to a level of $10\%$-$20\%$. 
The prospects are much more  open if the effects are exclusive to the first- and second-generation quark couplings, as the bounds on these are much more relaxed. NP could also show up in the Higgs self interaction. At the LHC, a direct probe of the Higgs self interaction is the di-Higgs production through gluon fusion, $gg\to hh$. At the leading order (LO), the self-coupling appears in a virtual-$h$-mediated process where the virtual Higgs is produced through quark triangles. There are also box diagrams, independent of the Higgs self coupling. In the SM, the di-Higgs cross section is almost three orders of magnitude smaller than the single Higgs production. This is because the box diagrams  interfere destructively with the triangle diagrams~\cite{Eboli:1987dy,Glover:1987nx,Plehn:1996wb,PhysRevD.58.115012,Djouadi:1999rca}, and producing two Higgses requires more energy than producing one (phase-space suppression). Ignoring the finite top-quark-mass effects, one gets $\sigma(gg\rightarrow hh)\approx 33$ fb with $\sqrt{s}=13$ TeV for $m_h=125$ GeV~\cite{deFlorian:2013jea,deFlorian:2015moa,Borowka:2016ehy,Borowka:2016ypz,DiMicco:2019ngk,Baglio:2018lrj,Baglio:2020wgt} at the next-to-next-to-leading order (NNLO) in $\alpha_s$. This accidental suppression of the di-Higgs production cross section makes the process a good candidate to probe for NP in the Higgs sector.

Significant progress has been made in the studies on di-Higgs production considering either the higher-order radiative corrections within the SM, or the presence of a NP candidate~\cite{PhysRevLett.111.201801,Frederix:2014hta, Maltoni:2014eza,GRIGO201417,
  Moretti_2005,PhysRevD.74.113008,Shao:2013bz,BARR2014308,
  BARGER2014433,deLima:2014dta,Arhrib_2009,Baglio:2012np,
  PhysRevD.87.055002,Han:2013sga,Goertz:2014qta, Dicus:2015yva,
  Delaunay:2013iia, PhysRevD.90.035016,Azatov:2015oxa,Contino:2012xk,
  Gillioz:2012se, Liu:2014rba,Christensen:2013dra,Gouzevitch:2013qca,
  Liu:2013woa, PhysRevD.89.095031,PhysRevD.90.015008,PhysRevD.90.055007,
  Hespel:2014sla,PhysRevLett.114.011801, Chen:2014ask,
vanBeekveld:2015tka,Ellwanger:2014hca,Hespel:2014sla,PhysRevD.60.075008,PhysRevD.64.035006,
  PhysRevD.82.115002, PhysRevD.86.095023,dawson:2015oha,Enkhbat:2013oba,
  Liu:2004pv, Dib:2005re, Chen:2014xra,
  Pierce_2007,PhysRevD.79.095010,HAN2010222,
  PhysRevD.87.014007,Edelhaeuser:2015zra,Lewis:2017dme,Lewis:2017dme,
  Huang:2017nnw,
  Alasfar:2019pmn,DaRold:2021pgn}. Prospects of several final states of the di-Higgs process, like the $bb\gamma\gamma$ \cite{Baur:2003gp,Baglio:2012np,Azatov:2015oxa,Kling:2016lay,Barger:2013jfa,Lu:2015jza,Adhikary:2017jtu,Alves:2017ued,Chang:2018uwu}, $bb\tau^+\tau^-$ \cite{Baglio:2012np,Adhikary:2017jtu,Goertz:2014qta,Dolan:2012rv}, and $bbbb$ \cite{Dolan:2012rv,deLima:2014dta,Behr:2015oqq,Wardrope:2014kya} channels have been studied. Various NP scenarios have been considered, e.g., the anomalous $t\bar{t}h$ ($b\bar{b}h$) and $t\bar{t}hh$ ($b\bar{b}hh$)
couplings~\cite{Delaunay:2013iia, PhysRevD.90.035016,Azatov:2015oxa,Contino:2012xk,Gillioz:2012se, Liu:2014rba}, resonant enhancements~\cite{Christensen:2013dra,Gouzevitch:2013qca,  Liu:2013woa, PhysRevD.89.095031,PhysRevD.90.015008,PhysRevD.90.055007,
  Hespel:2014sla,PhysRevLett.114.011801, Chen:2014ask,Abouabid:2021yvw},
new coloured scalars~\cite{Cao:2013si,Cao:2014kya,vanBeekveld:2015tka,Huang:2019bcs,
  PhysRevD.60.075008,PhysRevD.64.035006,PhysRevD.82.115002,
  Kribs:2012kz,dawson:2015oha,Enkhbat:2013oba,Huang:2017nnw,
  DaRold:2021pgn}, or fermionic particles~\cite{Liu:2004pv, Dib:2005re,
  Pierce_2007,PhysRevD.79.095010,HAN2010222,PhysRevD.87.014007,
  Edelhaeuser:2015zra} contributing to the loop amplitudes, etc. In a general effective-theory framework, one needs to consider the $qq\to hh$ process as well~\cite{Alasfar:2019pmn,Egana-Ugrinovic:2021uew}. 

Coloured bosons like Leptoquarks (LQs) can run in the loops of gluon initiated Higgs production~\cite{Agrawal:1999bk, Bhaskar:2020kdr} and, through large cubic and quartic interactions involving the Higgs, enhance the double Higgs production cross section~\cite{Enkhbat:2013oba,DaRold:2021pgn}. The high luminosity LHC (HL-LHC) can probe a LQ-induced enhancement of about twice the SM cross section with $\mathcal L\sim 2$ ab$^{-1}$ of integrated luminosity~\cite{DaRold:2021pgn}. However, as with any NP model, the enhancement depends on the allowed range of the LQ mass and its couplings to the SM particles. Collider phenomenology of various LQs has been extensively discussed in the literature~\cite{Dorsner:2016wpm,Mandal:2015lca,Das:2017kkm,Dorsner:2017ufx,Bandyopadhyay:2018syt, Hiller:2018wbv,Biswas:2018iak,Mandal:2018kau,Faber:2018afz,Alves:2018krf,Aydemir:2019ynb,Chandak:2019iwj,Padhan:2019dcp, Allanach:2019zfr,Bhaskar:2021pml,Bhaskar:2021gsy,Mandal:2015vfa}. 

In this letter, we extend this idea and consider the $pp\to hh$ process mediated by an $s$-channel SM-Singlet scalar resonance. Except for the Higgs, the singlet scalar couples to the SM fields only through its interaction with a scalar LQ. There are mainly three motivations for considering this setup. First, such a boson is well-motivated in many NP scenarios. Second, since,  unlike the Higgs, a singlet scalar cannot couple with the left-handed fermions, producing them at the colliders is not straightforward. Hence, studying the LHC phenomenology and prospects of a coloured-scalar-assisted production of a singlet scalar that dominantly decays to a Higgs pair is interesting. Third, the recent ATLAS searches for the resonant di-Higgs production ($pp\to X\to hh$) show a small excess of about $3.2\sigma$ local significance around $M_{X}\approx 1-1.1$ TeV~\cite{ATLAS:2021fet,ATLAS:2021tyg}. The analyses were done at $\sqrt s=13$ TeV with $139$ fb$^{-1}$ integrated luminosity for different final states: $b\bar bb\bar b$, $b\bar b\tau^+\tau^-$, $b\bar b\gamma\gamma$, and a combined one. Of these, the data in the $b\bar b\gamma\gamma$ mode go only up to a TeV. The enhancement is prominently visible in the $b\bar b\tau^+\tau^-$ and combined modes. The data put an upper bound on resonant $hh$ production cross section. The CMS collaboration has also put similar upper limits on the $hh$ cross section from various searches with $35.9$ fb$^{-1}$ and $138$ fb$^{-1}$ of data~\cite{CMS:2018ipl,CMS:2021roc}.\footnote{The excess is not visible in Ref.~\cite{CMS:2021roc} where only the leptonic decays of the $\tau$ lepton are considered.} Since the excess is not statistically significant, it could be just a statistical fluctuation. However, we take it as a motivation for choosing our model parameters to study the LHC phenomenology of the singlet scalar.

In principle, LQs can be either scalars or vectors in local quantum field theories. Here we consider a weak-singlet charge $-1/3$ scalar LQ (commonly known as $S_1$) for our purpose. The basic framework is the same as that in Ref.~\cite{Bhaskar:2020kdr} where we considered a simple extension of the SM augmented with an $S_1$ LQ and three generations of RH neutrino $\nu_R$. There we showed that the heavy neutrinos can induce a boost to the down-type-quark Yukawa interactions through a diagonal coupling with the quarks and $S_1$. The enhanced Yukawa couplings can also be realised  through a radiative generation of the dimension-$6$ operator of the form $ f_d (H^\dagger H/\Lambda^2)\left(\bar{q}_LH d_R\right) +{\rm H.c.}$ (with $\Lambda\sim$ TeV) which can enhance the \emph{single} Higgs production rate through the $qq\rightarrow h$ processes. In this framework, the coefficient $f_d$ is restricted by LHC limits on the masses and the couplings of new fields (e.g. LQs, $\nu_R$) to the SM ones. However, the limits become weaker, if the SM Higgs boson is replaced by a \emph{new}  scalar $\phi$. Hence, a large production rate can be obtained for the $qq\rightarrow \phi$ processes. Through the same process, a  resonant $\phi$ can generate the $hh$ final state at the LHC, which is further enhanced at the resonance. We will see if any significant excess in the $2h$ cross-section can be observed if the gluon fusion process is supplemented by the quark fusion. We will also investigate the prospects of this process at the HL-LHC.

The plan of the letter is as follows. In the next section, we review the model with the $S_1$ LQ; in Sec.~\ref{sec:pp2hh}, we discuss the di-Higgs production in the SM and via the new scalar resonance; in Sec.~\ref{sec:pheno}, we discuss the prospects of the channel at the HL-LHC and  conclude in Sec~\ref{sec:conclu}.

\section{A recap of the LQ Model}\label{sec:model}
\noindent
The model in Ref.~\cite{Bhaskar:2020kdr} is an extension of the SM with chiral neutrinos and an $S_1$-type scalar LQ. The LQ transforms under the SM gauge group as $\displaystyle\left ({\bf\bar{3}},{\bf 1},1/3\right)$ with $Q_{\rm EM}=T_3+Y$. For our purpose, we introduce a real SM singlet scalar $\phi$. The general fermionic interaction Lagrangian for $S_1$ in addition to the kinetic term can be written in the notation of Ref.~\cite{Dorsner:2016wpm} as, 
\begin{align}
\nonumber	\mathcal{L^\prime} = &\ (D_\mu S_1)^\dagger(D^\mu S_1)+(\partial_\mu \phi)^\dagger(\partial^\mu \phi)\nonumber\\
&-\Big[(y_1^{LL})_{ij} (\bar{Q}_L^{Cia}\epsilon^{ab}L_L^{jb})S_1+ (y_1^{RR})_{ij} 
(\bar{u}_R^{Ci}e_R^j)S_1\nonumber\\
&\quad + (y_1^{\overline{RR}})_{ij} (\bar{d}_R^{Ci}\nu_R^j)S_1+{\rm H.c.}\Big],
\label{eq:lagrangianF}
\end{align}
where $D_\mu=\partial_\mu+ig_ST^aG^a_\mu+iYB_\mu$ is the covariant derivative; $T^a$ and $G^a_\mu$ are the colour generators and the gluon fields, respectively; $B_\mu$ is the  gauge mediator of  hypercharge $Y$ and $g_S$ is the strong coupling. We have suppressed the colour indices. The superscript $C$ denotes charge conjugation; $\{i,j\}$ and $\{a,b\}$ are flavour and $SU (2)$ indices, respectively. The SM quark and lepton doublets are denoted by $Q_L$ and $L_L$, respectively. We add the scalar interaction terms to the Lagrangian in Eq.~\eqref{eq:lagrangianF} as,
\begin{align}
\mathcal{L}\supset\ \mathcal L^\prime&\ -\Big[\lambda \left(H^\dagger H\right) \left(S^\dag_1S_1\right)+ \lambda^\prime  \phi\left(S^\dag_1S_1\right)+\mu\left(H^\dagger H\right)\phi^2\nonumber\\
&\quad +\mu^\prime\left(H^\dagger H\right)\phi+\frac{1}{2}M_\phi^2\phi^2+\bar M^{2}_{S_1}\left(S^\dag_1S_1\right)\Big].
\label{eq:lagrangian}
\end{align}
Here, $H$ denotes the SM Higgs doublet, and $\bar M_{S_1}$ and $M_\phi$ define the bare mass parameters for $S_1$ and $\phi$ respectively, and $\lambda^{\prime}$ and $\mu^{\prime}$ are  dimension-one couplings proportional to some NP scale. For simplicity, we shall assume $\mu\rightarrow 0$. We denote the physical Higgs field after the electroweak symmetry breaking as $h\equiv h_{125}$. The physical masses can be obtained via
\begin{equation}
H = 
\frac{1}{\sqrt{2}}\begin{pmatrix}
0 \\
v + h
\end{pmatrix},
\end{equation}
where $v \simeq 246~$GeV is the vacuum expectation value (VEV) of the SM Higgs. The physical mass of $S_1$ is given as
\begin{equation}
M^2_{S_1}~=~\bar M^{2}_{S_1}+ \frac{1}{2}\lambda v^2.
\end{equation}

In the subsequent discussion, we set the quark and neutrino mixing matrices to identity, assume the LQ couplings to be flavour diagonal, and put $y_1^{RR}=0$ to avoid flavour bounds~\cite{Mandal:2019gff}. We write $y_1^{LL}=g_L$ and $y_1^{\overline{RR}}=g_R$ for simplicity. For a closer look, we can expand Eq.~\eqref{eq:lagrangian} for the first generation as
	\begin{align}\label{eq:Lag}
	\mathcal{L}_F\ \supset&\ (D_\mu S_1)^\dagger(D^\mu S_1)+(\partial_\mu \phi)^\dagger(\partial^\mu \phi)\nonumber\\
	&\ -\Big[g_L\left(-\bar{d}_L^C\nu_L+ \bar{u}_L^C e_L\right)S_1
+ g_R\ \bar{d}_R^C \nu_R S_1 + H.c.\Big],
	\end{align}
where we have simplified $\displaystyle \left(y_1^{X}\right)_{ii}$ as $y_i^{X}$. Since the flavour of the neutrino is irrelevant for the LHC, we simply write $\nu$ to denote a neutrino. The LQ-gluon couplings are generated from the covariant derivative term:
\begin{align}
\mathcal{L}_{F}\supset&\ \Big[(\partial_\mu+ig_S \mathcal{G}_\mu)S_1\Big]^\dagger\Big[(\partial^\mu+ig_S \mathcal{G}^{\mu})S_1\Big]\nonumber\\
=&\ (\partial_\mu S_1)^\dagger(\partial^\mu S_1)-ig_S\Big[(\mathcal{G}_{\mu}S_1)^\dagger(\partial^\mu S_1)\nonumber\\
&\quad-(\partial_\mu S_1)^\dagger(\mathcal{G}^{\mu}S_1)\Big]+g_S^2(\mathcal{G}_{\mu}S_1)^\dagger(\mathcal{G}^{\mu}S_1)
\label{eq:DS}
\end{align}
where, $\mathcal{G}_\mu=T^aG^a_\mu$. The second term of Eq.~\eqref{eq:DS} produces the $gS_1S_1$ vertex~[vertex factor $\sim -g_S(p+p^\prime)^\mu$] while the last term is responsible for $ggS_1S_1$ four point interaction~[vertex factor $\sim g_S^2$]. Here $p$ and $p^\prime$ are the incoming and outgoing momenta of the of $S_1$.

\begin{figure}[!t]
\captionsetup[subfigure]{labelformat=empty}
\centering
\subfloat{\includegraphics[width=0.9\columnwidth]{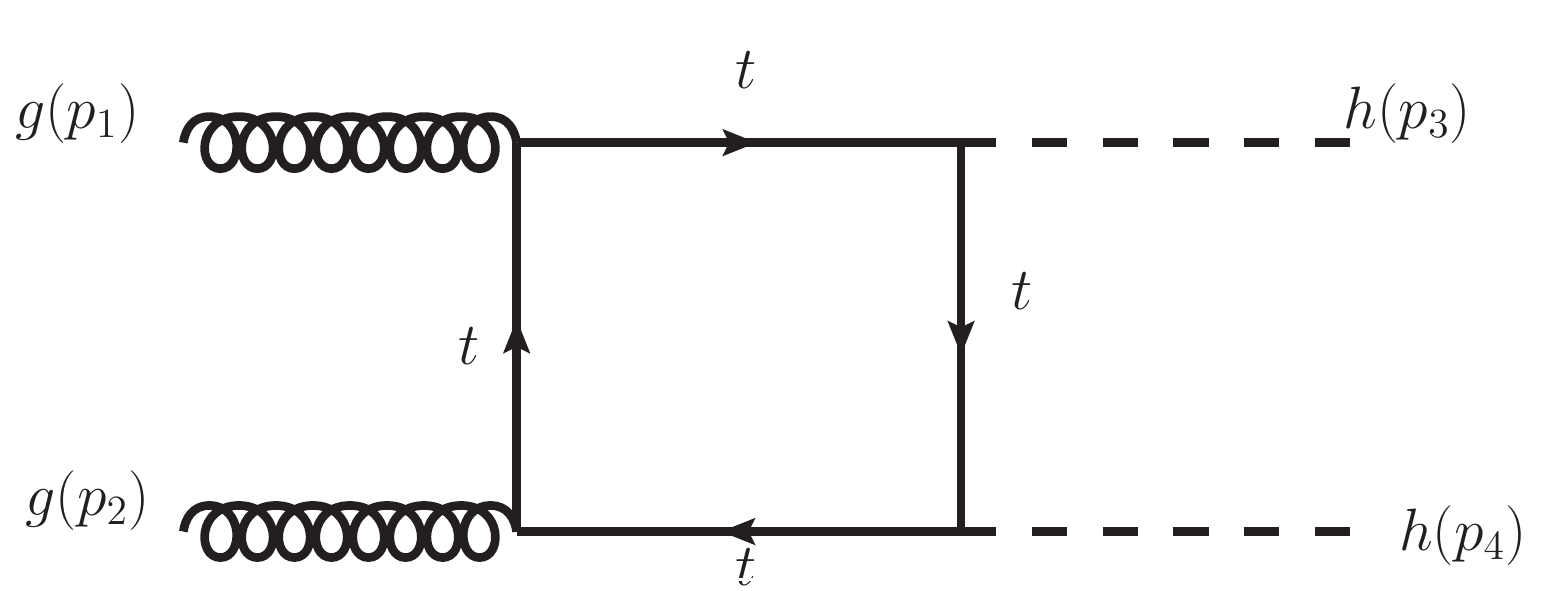}\label{fig:gghh_SM1}}\\
\subfloat{\includegraphics[width=0.9\columnwidth]{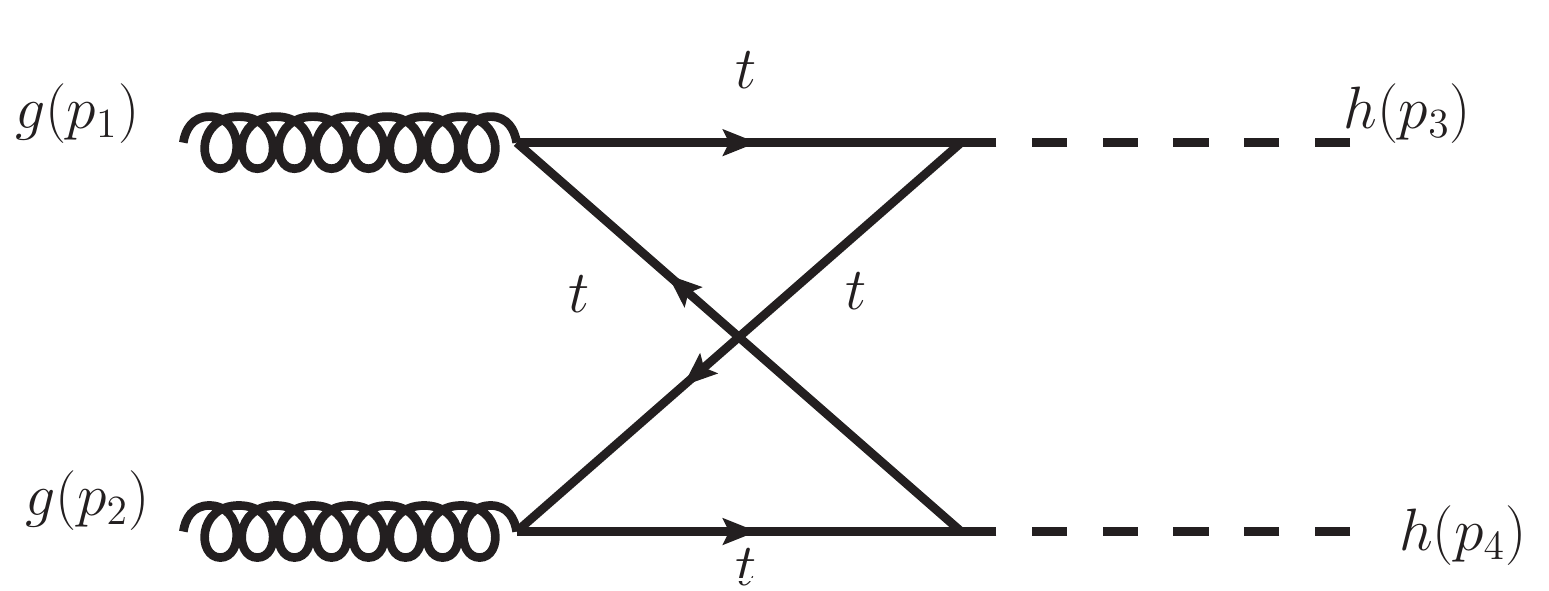}\label{fig:gghh_SM2}}\\
\subfloat{\includegraphics[width=0.9\columnwidth]{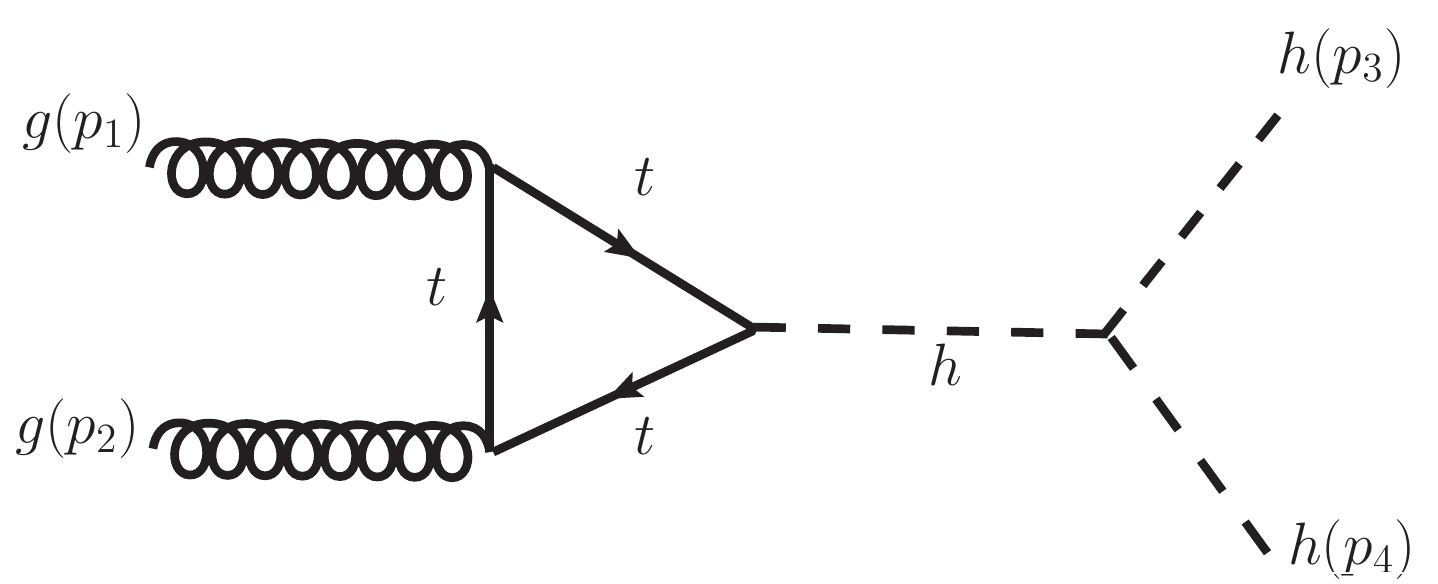}\label{fig:gghh_SM3}}
\caption{Feynman diagrams for $gg \rightarrow hh$ in the Standard Model.}
\label{fig:gghh_SM}
\end{figure}

\subsection{Mixing between $\phi$ and the SM Higgs}
\noindent 
After the electroweak symmetry breaking, $h$ and $\phi$ mix through the term, $\mu^\prime\left(H^\dagger H\right)\phi$. In the $(\phi\quad h)^T$ basis, the mass matrix can be written as,
\begin{align}
M_{\phi h}=\left(\begin{array}{c c}
M_\phi^2 & \mu^\prime v\\
\mu^\prime v & m_h^2
\end{array}\right).
\label{eq:Mph}
\end{align}
We can rotate $(\phi\quad h)^T$ to a physical basis $(H_1\quad H_2)^T$ as:
\begin{align}
\left(\begin{array}{c}
\phi\\
h
\end{array}\right)=\left(\begin{array}{c c}
\cos\theta & \sin\theta\\
-\sin\theta & \cos\theta
\end{array}\right)\left(\begin{array}{c}
H_1\\
H_2
\end{array}\right).
\label{eq:U}
\end{align}
In the new basis, the  mass matrix is diagonal and it is defined as, 
\begin{align}
\mathcal{M}_D=U^\dagger M_{\phi h}U=\left(\begin{array}{c c}
M^2_{H_1} & 0\\
0 & M^2_{H_2}
\end{array}\right),
\end{align}
where $U$ is an orthogonal rotation matrix and 
\begin{align}
M^2_{H_1}&=M_\phi^2\cos^2\theta+m_h^2\sin^2\theta-\mu^\prime v\sin 2\theta,\\
M^2_{H_2}&=M_\phi^2\sin^2\theta+m_h^2\cos^2\theta+\mu^\prime v\sin 2\theta.
\end{align}
The mixing angle is given as,
\begin{align}
\tan 2\theta=\frac{2\mu^\prime v}{m_h^2-M_\phi^2}.
\end{align}
For a small mixing angle, we can assume $H_1$ to be mostly singlet-like while $H_2$ will show a strong doublet-like nature. Using Eq.~\eqref{eq:U}, we can recast the important interaction terms as,
\begin{align}
\lambda v h\left(S^\dag_1S_1\right)=&\ \lambda v(-H_1s_\theta+H_2c_\theta)\left(S^\dag_1S_1\right)\\
\lambda^\prime  \phi\left(S^\dag_1S_1\right)=&\ \lambda^\prime  (H_1c_\theta+H_2s_\theta)\left(S^\dag_1S_1\right)\\
\frac{\mu^\prime}{2}h^2\phi=&\ \frac{\mu^\prime}{2}\left(H_1H_2^2c^3_\theta+H_2^3c^2_\theta s_\theta\right.\nonumber\\ 
&\quad\quad\left.-2H_1H_2^2 s^2_\theta c_\theta+\cdots\right)
\end{align}
where $s_\theta=\sin\theta$ and $c_\theta=\cos\theta$.

In the subsequent analysis, we assume $c_\theta \to 1$ and $s_\theta \to 0$. In this limit, we assign $m_{h_{SM}} \simeq M_{H_2}=m_h$ and $M_\phi = M_{H_1}$ in the following sections. The above assignment is valid only if $\mu^\prime \to 0$ or 
$M_\phi \gg m_h>\mu^\prime$. In our case, the second condition is applicable as we are primarily interested in a  new scalar resonance at $\sim 1.1$ TeV in the di-Higgs production rate. 

\section{Di-Higgs production through a scalar resonance in the LQ model}\label{sec:pp2hh}
\noindent
We first look at the leading-order (LO) SM contributions to the di-Higgs production at the LHC.
\subsection{Di-Higgs production in the SM}
\noindent
Here we start with reviewing the dominant production processes for the $gg\to hh$ ~\cite{Eboli:1987dy,Glover:1987nx,Plehn:1996wb,PhysRevD.58.115012,Djouadi:1999rca}, as shown in Fig.~\ref{fig:gghh_SM}. The amplitude for $g^{a\mu}(p_1)g^{b\nu}(p_2)\rightarrow h(p_3)h(p_4)$ is given by,
\begin{align}
    \mathcal{A}_{ab}^{\mu\nu}=\ \frac{\alpha_S\delta_{ab}}{8\pi v^2}&\Big[\mathcal{P}^{\mu\nu}(p_1,p_2)\mathcal{F}(\hat{s},\hat{t},\hat{u},m_t^2)\nonumber\\
    &\ +\mathcal{Q}^{\mu\nu}(p_1,p_2,p_3)\mathcal{G}(\hat{s},\hat{t},\hat{u},m_t^2)\Big],
\end{align}
where $a,b$ are the colour indices, $\hat{s},\hat{t},\hat{u}$ are the parton-level Mandelstam variables. The independent tensor projections $\mathcal{P}^{\mu\nu}$ and $\mathcal{Q}^{\mu\nu}$ are defined as,
\begin{align}
    \mathcal{P}^{\mu\nu}=&\ g^{\mu\nu}-\frac{p_1^\nu p_2^\mu}{p_1.p_2},\\
    \mathcal{Q}^{\mu\nu}=&\ g^{\mu\nu}+\frac{2}{\hat{s}p_T^2}\Big[m_h^2p_1^\nu p_2^\mu - 2(p_1.p_3)p_2^\mu p_3^\nu\nonumber\\
    &\hspace{1.7cm}- 2(p_2.p_3)p_1^\nu p_3^\mu+\hat{s}p_3^\mu p_3^\nu\Big],
\end{align}
where $p_T=(\hat{u}\hat{t}-m_h^4)/\hat{s}$ is the transverse momentum of an outgoing Higgs. The triangle diagram contributes only to $\mathcal{F}(\hat{s},\hat{t},\hat{u},m_t^2)$, while box diagrams contribute to both $\mathcal{F}(\hat{s},\hat{t},\hat{u},m_t^2)$ and $\mathcal{G}(\hat{s},\hat{t},\hat{u},m_t^2)$. The triangle diagram has no angular momentum dependence, thus, it is an s-wave contribution. The form factors $\mathcal{F}^{\rm box}$ and $\mathcal{G}^{\rm box}$ can be attributed to the spin-$0$ and spin-$2$ parts of the box amplitude, respectively. The angular dependencies of the form factors by considering the partial wave decomposition of the scattering amplitude are discussed in Refs.~\cite{Dawson:2012mk,Borowka:2016ypz}. We can split the SM contribution to $\mathcal{F}$ as $\mathcal{F}_{\rm SM}=\mathcal{F}^{\rm tri}_{\rm SM}+\mathcal{F}^{\rm box}_{\rm SM}$ where
\begin{align}
\mathcal{F}^{\rm tri}_{\rm SM}=&\frac{12m_h^2m_t^2}{\hat{s}-m_h^2}\Big[2+(4m_t^2-\hat{s})C_0(0,0,\hat{s},m_t^2,m_t^2,m_t^2)\Big] \label{Eq:SMtriangle}\\
\mathcal{F}^{\rm box}_{\rm SM}=&-\frac{2m_t^2}{\hat{s}}\Bigg[-4\hat{s}-8m_t^2\hat{s}C^{00}_{ttt}(\hat{s})-2(4m_t^2-m_h^2)\nonumber\\
&\quad\times\Big\{2(m_h^2-\hat{t})C^{h0}_{ttt}(\hat{t})+2(m_h^2-\hat{u})C^{h0}_{ttt}(\hat{u})\nonumber\\
&\quad-(m_h^4-\hat{t}\hat{u})D^{h0h0}_{tttt}(\hat{t},\hat{u})\Big\}+2m_t^2\hat{s}(2m_h^2-8m_t^2+\hat{s})\nonumber\\
&\quad\times\Big\{D^{hh00}_{tttt}(\hat{s},\hat{t})+D^{hh00}_{tttt}(\hat{s},\hat{u})+D^{h0h0}_{tttt}(\hat{t},\hat{u})\Big\}\Bigg].
\end{align}
Here we have used the following short hands for three-point and four-point functions,
\begin{align}
C^{ab}_{ijk}(\hat{z}) =& C_0 (m^2_a , m^2_b, \hat{z}, m^2_i ,m^2_j , m^2_k)\quad {\rm and}\nonumber\\
D^{abcd}_{ijkl}(\hat{w},\hat{z})=&D_0(m_a^2,m_b^2,m_c^2,m_d^2,\hat{w},\hat{z},m_i^2,m_j^2,m_k^2,m_l^2).
\end{align}
We have ignored all other quark loops except the $t$-quark loop since it gives the dominant contribution. Similarly, $\mathcal{G}_{\rm SM}(\hat{s},\hat{t},\hat{u},m_t^2)$ can be expressed as,
\begin{align}
\mathcal{G}_{\rm SM}=&\frac{2m_t^2}{m_h^4-\hat{t}\hat{u}}\Bigg[(2m_h^4-\hat{t}^2-\hat{u}^2)(8m_t^2-\hat{t}\hat{u})C^{hh}_{ttt}(\hat{s})\nonumber\\
&\quad+(m_h^4-8m_t^2\hat{t}+\hat{t}^2)\Big\{2(m_h^2-\hat{t})C^{0h}_{ttt}(\hat{t})-\hat{s}C^{00}_{ttt}(\hat{s})\nonumber\\
&\quad+\hat{s}\hat{t}D^{00hh}_{tttt}(\hat{s},\hat{t})\Big\}+(m_h^4-8m_t^2\hat{u}+\hat{u}^2)\nonumber\\
&\quad\times\Big\{2(m_h^2-\hat{u})C^{0h}_{ttt}(\hat{u})-\hat{s}C^{00}_{ttt}(\hat{s})+\hat{s}\hat{u}D^{00hh}_{tttt}(\hat{s},\hat{u})\Big\}\nonumber\\
&\quad+2m_t^2(m_h^4-\hat{t}\hat{u})(8m_t^2-\hat{t}-\hat{u})\Big\{D^{0h0h}_{tttt}(\hat{t},\hat{u})\nonumber\\
&\quad+D^{00hh}_{tttt}(\hat{s},\hat{t})+D^{00hh}_{tttt}(\hat{s},\hat{u})\Big\}\Bigg].
\end{align}
As shown, the leading order (LO) amplitude includes the top-quark-mass-dependent terms. The contribution at the next-to-leading order (NLO) in the strong coupling constant both in the infinite top-mass limit and with the full top-mass dependence are considered in Refs.~\cite{Dawson:1998py,Alasfar:2019pmn,Borowka:2016ehy,Borowka:2016ypz,Baglio:2018lrj,Baglio:2020wgt}. For higher-order results
in different top mass limits, see Refs.~\cite{Grigo:2014jma,deFlorian:2013jea,Grazzini:2018bsd,Davies:2019nhm,Davies:2019djw}. 

\begin{figure}[!t]
\captionsetup[subfigure]{labelformat=empty}
\begin{center}
\subfloat[\quad\quad(a)]{\includegraphics[width=0.9\columnwidth]{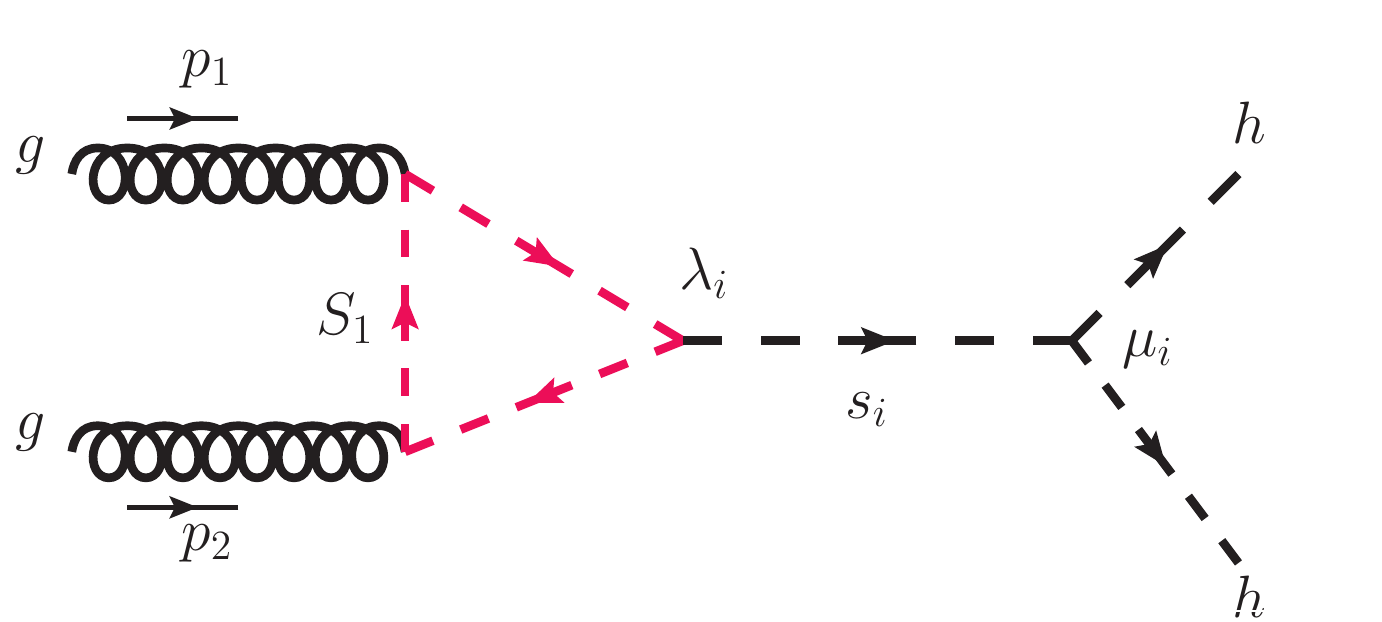}\label{fig:gghh_r1}}\\
\subfloat[\quad\quad(b)]{\includegraphics[width=0.9\columnwidth]{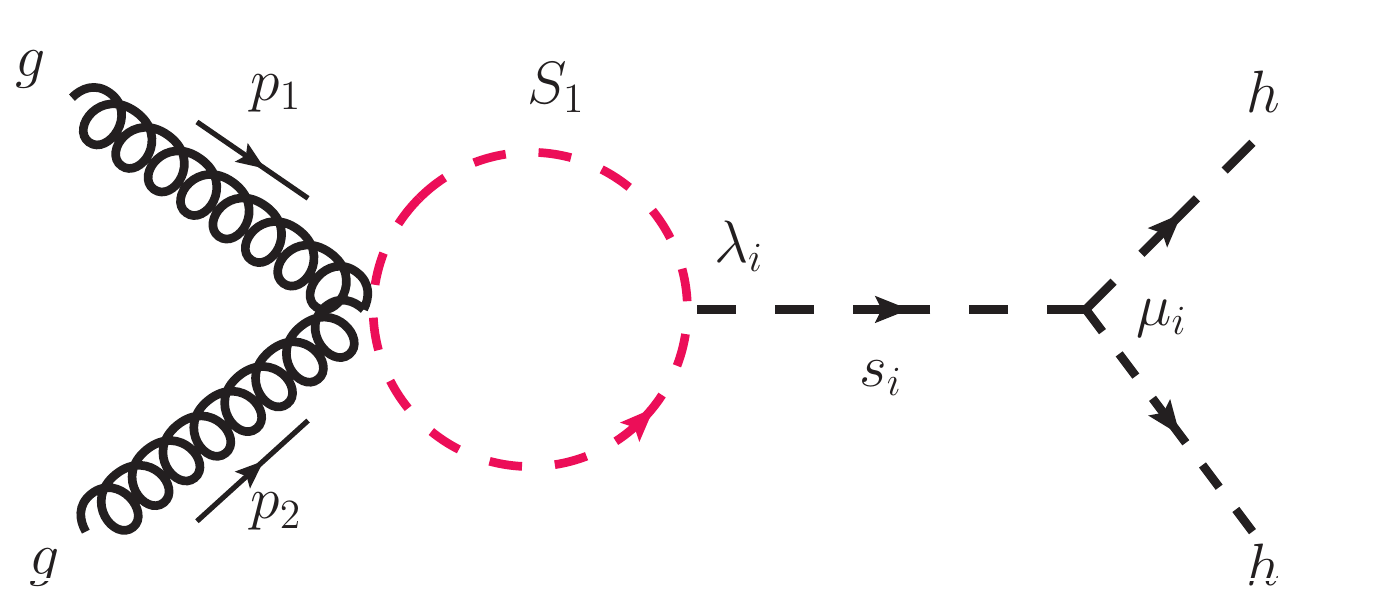}\label{fig:gghh_r2}}
\end{center}
\caption{Resonance diagrams contributing to $\sigma(gg\rightarrow hh)$.}
\label{fig:gghh_r}
\end{figure}
\subsection{The $gg\rightarrow hh$ process in presence of $\phi$}\label{subsec:gghh}
\noindent
The  diagrams for the $gg\rightarrow hh$ through the exchanges of a scalar $s_i$ where $s_i=\phi,\, h$ are shown in Fig.~\ref{fig:gghh_r}. The complete invariant amplitude for the $gg\rightarrow hh$  processes can be obtained by considering the contributions from both the triangle~[Fig.~\ref{fig:gghh_r1}] and  bubble~[Fig.~\ref{fig:gghh_r2}] diagrams:
\begin{align}
\mathcal{A}^{s_i}_{gghh}= \frac{\alpha_S\delta_{ab}}{8\pi v^2}\mathcal{P}^{\mu\nu}(p_1,p_2)\mathcal{F}^{s_i}_{gghh}.
\end{align}

The loop function in the $gg\,s_i$ effective vertex is given by,
\begin{align}
\mathcal{F}^{s_i}_{gghh}=-\frac{\lambda_i\mu_iv^2}{\hat{s}-M_{s_i}^2}\Big[1+2M_{S_1}^2C_0(0,0,\hat{s},M_{S_1}^2,M_{S_1}^2,M_{S_1}^2)\Big].\label{eq:propgg}
\end{align}
For the SM Higgs, $\lambda_i=\lambda v$ and $\mu_i=3m_h^2/v$, while for the singlet scalar $\phi$, $\lambda_i=\lambda^\prime$ and $\mu_i=\mu^\prime/2$. The total $gg\rightarrow hh$ differential cross section including all the resonance channel diagrams~(SM+NP) and the box diagrams~(SM only) can be calculated as,
\begin{align}
\frac{d}{d\hat{t}}\hat{\sigma}(gg\rightarrow hh)=\frac{\alpha_S^2G_F^2}{2^{14}\pi^3\hat{s}^2}\Bigg(\left|\mathcal{F}+\sum_{s_i=h,\phi}\mathcal{F}^{s_i}_{gghh}\right|^2+\left|\mathcal{G}\right|^2\Bigg).
\label{eq:sig_gg}
\end{align}

\begin{figure}[!t]
\captionsetup[subfigure]{labelformat=empty}
\begin{center}
\subfloat[\quad\quad(a)]{\includegraphics[scale=0.6]{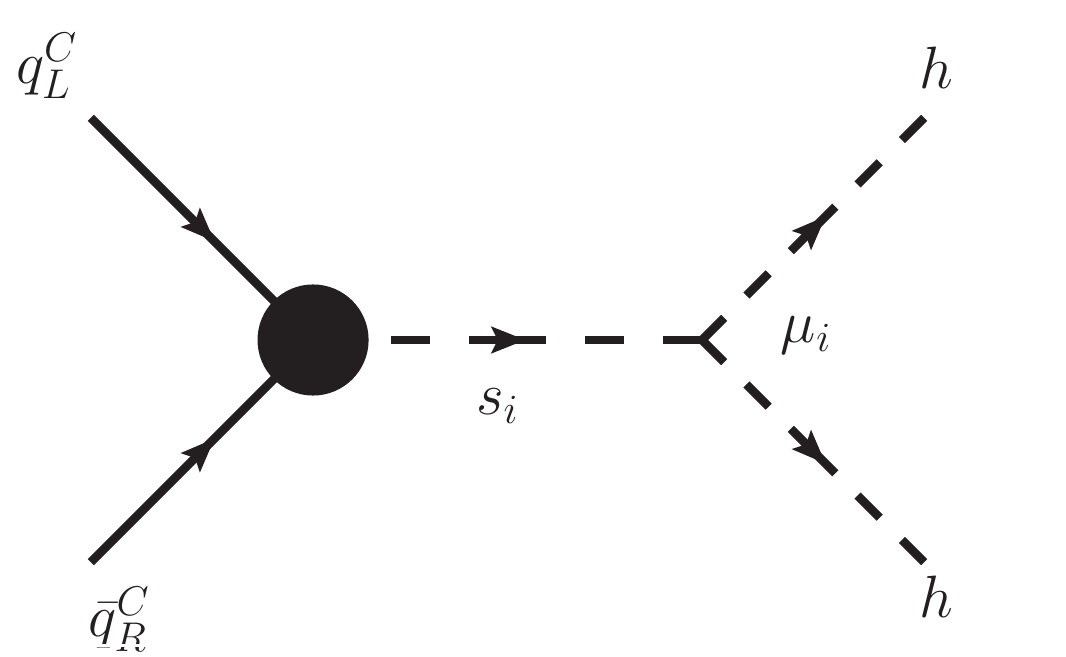}\label{fig:qqhh}}\\
\subfloat[\quad\quad(b)]{\includegraphics[scale=0.5]{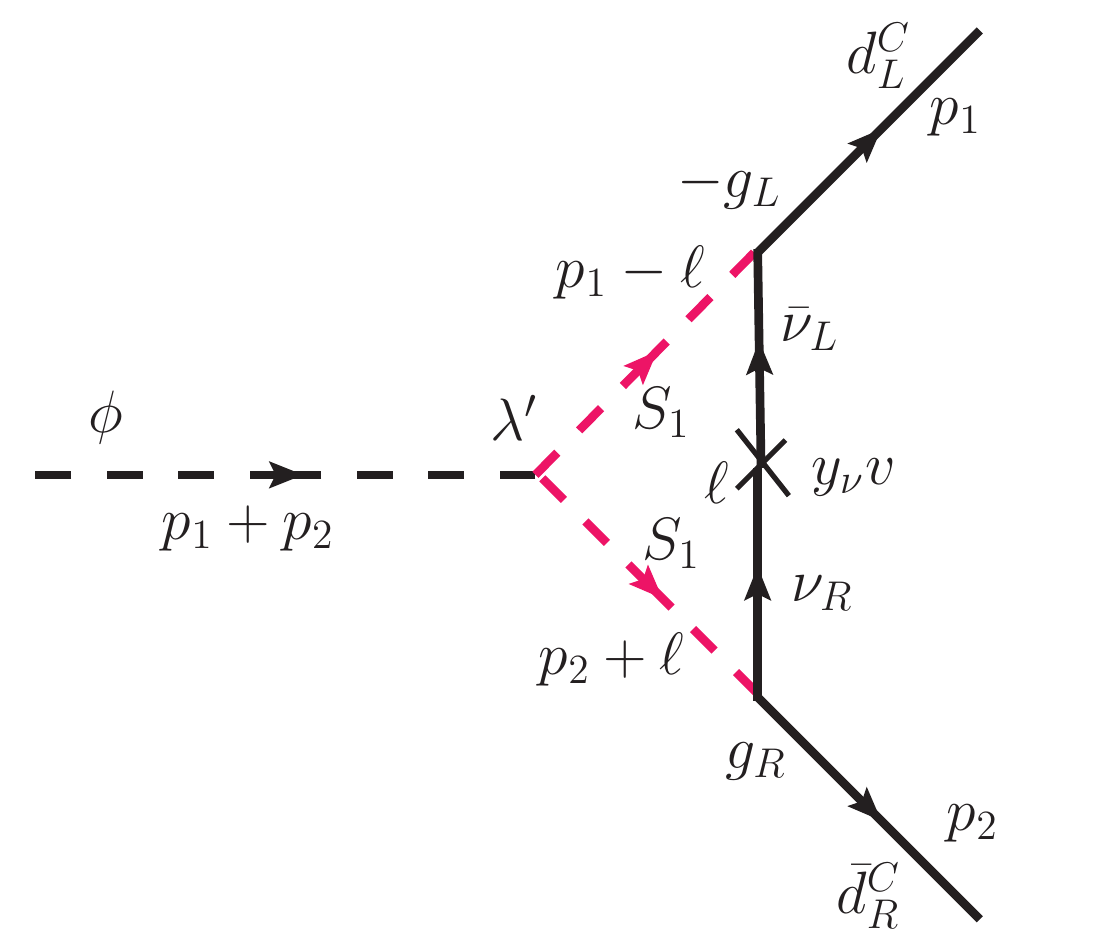}\label{fig:qqhh_1}}\\
\end{center}
\caption{(a) Representative Feynman diagram for the $qq\rightarrow hh$ processes and (b) the $q\bar{q}\phi$ vertex.}
\label{fig:qqfusion}
\end{figure}
\subsection{The ${qq\rightarrow hh}$ process}
\noindent
Fig.~\ref{fig:qqhh} shows a generic diagram for the $qq\rightarrow hh$ processes, where $s_i=\{\phi,\, h\}$ and $q$ is a down-type quark. The blob represents the $s_i q\bar q$ effective vertex. The effective couplings are available in Ref.~\cite{Bhaskar:2020kdr} which we repeat below for completeness.
\begin{itemize}
\item {\bfseries For $s_i=\phi$:} For the singlet scalar, there is only one possible vertex, as shown in Fig.~\ref{fig:qqhh_1}. The corresponding effective coupling is given by,
\begin{align}
Y_{q\bar{q}\phi}=-\frac{\mathcal C_q\lambda^\prime  v}{16\pi^2}\,C_0(0,0,\hat{s},M_{\nu_R}^2,M_{S_1}^2,M_{S_1}^2),
\end{align}
where we use a compact notation, $\mathcal C_q=g_Lg_Ry_\nu$.
\item {\bf For $s_i=h$:} For the SM Higgs, there are three possible one loop diagrams along with the tree-level contribution \cite{Bhaskar:2020kdr}.  The total effective coupling for the $q\bar{q}h$ vertex is given by,
\begin{align}
Y_{q\bar{q}h}\ =&\ \ {y^{\rm SM}_q}+\frac{\mathcal C_q}{16\pi^2}
\Big[B_0 (0,M_{\nu_R}^2,M_{S_1}^2)-B_0 (\hat{s},0,M_{\nu_R}^2)\nonumber\\
&\ - M_{S_1}^2C_0 (0,0,\hat{s},0,M_{S_1}^2,M_{\nu_R}^2)\nonumber\\
&\ - \lambda v^2 C_0 (0,0,\hat{s},M_{S_1}^2,M_{\nu_R}^2,M_{S_1}^2)\nonumber\\
&\ +\frac{\lambda v^2}{2}\Big\{C_0(0,0,\hat{s},0,M_{S_1}^2,M_{\nu_R}^2) \nonumber\\
&\ + M_{S_1}^2 D_0(0,0,\hat{s},0,0,0,M_{S_1}^2,M_{S_1}^2,0,M_{\nu_R}^2)\nonumber\\
&\ - C_0(0,0,0,M_{S_1}^2,M_{S_1}^2,M_{\nu_R}^2)\Big\} \Big],
\label{eq:MtotPV}
\end{align}
where $D_0$, $C_0$ and $B_0$ are the Passarino-Veltman four-point, triangle, and bubble integrals, respectively. The ${y^{\rm SM}_q}$ term is the SM tree-level contribution. 
\end{itemize}
The differential cross section for $q\bar{q}\rightarrow hh$ at the leading order is given by,
\begin{align}
    \frac{d\hat{\sigma}(q\bar{q}\rightarrow hh)}{dt}=\frac{1}{16\pi}\frac{1}{12\hat{s}}\left|\sum_{s_i=h,\phi}\frac{\mu_iY_{q\bar{q}s_i}}{\hat{s}-M_{s_i}^2}\right|^2.\label{eq:propqq}
\end{align}

%
\begin{figure}[!t]
\captionsetup[subfigure]{labelformat=empty}
\begin{center}
\subfloat{\includegraphics[width=0.9\columnwidth]{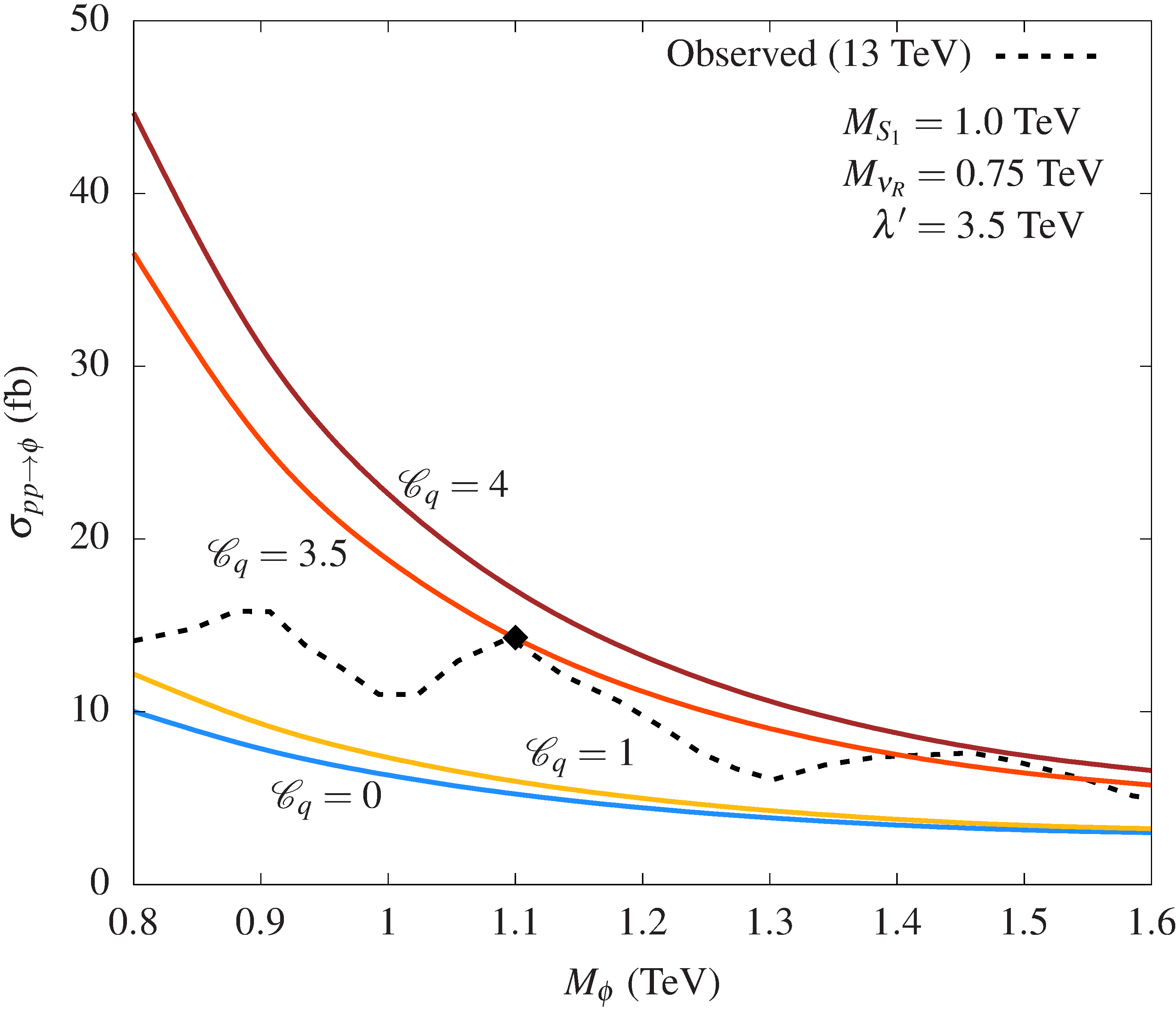}}\\
\end{center}
\caption{The $pp\rightarrow \phi$ cross section for different values of $\mathcal{C}_q$  as functions of $M_{\phi}$ for $\sqrt{s}= 13$ TeV. The combination $\mathcal{C}_q= g_L g_R y_{\nu}$ parametrises the quark contribution. To obtain the cross sections we have used the $14$ TeV di-Higgs K-factors, K$_{qq~(gg)}= 1.3~(1.72)$, from Ref.~\cite{Alasfar:2019pmn}. Since, BR($\phi\to hh)\approx 1$ in our model, $\sigma(pp\rightarrow \phi)\approx\sigma(pp\rightarrow \phi\to hh)$ [see Eq.~\eqref{eq:narroww}]. Hence, we compare it with the  upper limit on $\sigma(X\to hh)$ (dashed line) as observed by the ATLAS collaboration in the combined channel~\cite{ATLAS:2021tyg}. The black diamond signifies the parameter choice for which $\sigma(pp\rightarrow \phi)\approx 14.5$ fb at $M_\phi=1.1$ TeV.}
\label{fig:CSMphi}
\end{figure}
\subsection{The resonant $\mathbb \phi$ contribution and choice of parameters}
\noindent

By replacing the $s_i$ propagator in Eqs.~\eqref{eq:propgg} and \eqref{eq:propqq} by the Breit-Wigner shape,
\begin{align}
    \frac{1}{\hat s - M_{s_i}^2} \to \frac{1}{\hat s - M_{s_i}^2 +iM_{s_i}\Gamma_{s_i}},
\end{align}
we obtain the gluon and quark contributions to the di-Higgs process including the SM-NP interference. However, since we focus on a TeV-range $\phi$, the $\phi$ resonance appears far from the Higgs contribution, making the interference negligible. Hence, in this case, it is possible to look at the resonant $\phi$ contribution separately. This is also justified by the fact that in the parameter range of our interest, the decay width of $\phi$ is quite narrow, $\Gamma_{\phi}/M_\phi < 1\%$ (for example, for $M_{S_1}=1$ TeV, $M_{\nu_R}=0.75$ TeV, $\mathcal{C}_{q}=3.5$, $\lambda^{\prime}=3.5$ TeV and $\mu^\prime = 200$ GeV, we get $\Gamma_{\phi}\lesssim 1$ GeV). As a result, we could use the narrow-width approximation to estimate the resonant $\phi$ contribution to the di-Higgs production. In other words,
\begin{align}
    \sigma^{qq(gg)\to\phi\to hh}_{\rm had}= \sigma^{qq(gg)\rightarrow \phi}_{\rm had}\times {\rm BR}(\phi \rightarrow h h),
\end{align} 
where BR stands for branching ratio.
For $\mu^\prime\gtrsim 100$ GeV, the tree-level decay of $\phi$ dominates, making BR$(\phi \to hh)\approx 1$. We take a conservative $\mu^\prime= 200$ GeV to keep $\theta$ small. Hence, we can further simplify the above expression as
\begin{align}
    \sigma^{qq(gg)\to\phi\to hh}_{\rm had}\approx \sigma^{qq(gg)\rightarrow \phi}_{\rm had}.\label{eq:narroww}
\end{align}

We show $\sigma(pp\to\phi)$ at the $\sqrt{s}= 13$ TeV LHC in Fig.~\ref{fig:CSMphi}. The combination $\mathcal{C}_q=g_Lg_Ry_\nu$ parametrises the quark-initiated contribution. The quark contribution switches off for $\mathcal{C}_q=0$. Since $\sigma^{qq}$ is proportional to $\mathcal C^2_q$, $\sigma(pp\to\phi)$ increases rapidly with increasing $\mathcal C_q$. For this plot, we use a benchmark set of parameters, $M_{\nu_R}=0.75$ TeV, $M_{S_1}=1$ TeV and $\lambda^\prime=3.5$ TeV. Our choice of the right-handed neutrino mass is motivated by Ref.~\cite{ATLAS:2020syg,ATLAS:2021yij}. The current direct search limits on $S_1$ lie between $1.7-1.8$ TeV~\cite{ATLAS:2020dsk}. These limits are obtained assuming that the $S_1$ decays exclusively to one or two final states (like, e.g., $ej$ or $ej$ and $\nu_ej$). However, in our case, the $S_1$ can decay to $ue,c\mu,t\tau,d\nu,s\nu,b\nu,d\nu_R,s\nu_R,s\nu_R$ final states, making the BR in each light  mode (i.e., those without $\nu_R$---there  is no direct limit available from these exotic decay modes) to be less than $1/6$~\cite{Bhaskar:2020kdr}. As a result, a TeV $S_1$ is allowed. Moreover, the combination $\mathcal{C}_q=g_Lg_Ry_\nu$ can be order one even when $g_L$ is less than one. Thus, it is possible to take $M_{S_1}= 1$ TeV and still avoid the constraints from low-energy observables and the single-production searches. Finally, the dimension-one coupling $\lambda^\prime$ comes from a NP scale which we  assume to be heavier than the entire particle spectrum of our model. The particular choice is motivated by the fact that for $\lambda^\prime\sim 3.5$ TeV, $\sigma(pp\to\phi)$ at $13$ TeV fits the small excess seen by ATLAS~\cite{ATLAS:2021tyg} at $M_\phi=1.1$ TeV with perturbative couplings.

\begin{table}[t!]
\centering{\linespread{2}
\begin{tabular*}{\columnwidth}{l @{\extracolsep{\fill}} crc }
\hline
\multicolumn{2}{l}{Background } & $\sigma$ & QCD\\ 
\multicolumn{2}{l}{processes}&(pb)&order\\\hline\hline
\multirow{2}{*}{$V +$ jets~ \cite{Catani:2009sm,Balossini:2009sa}  } & $Z +$ jets  &  $6.33 \times 10^4$& NNLO \\ 
                & $W +$ jets  & $1.95 \times 10^5$& NLO \\ \hline
\multirow{3}{*}{$VV +$ jets~\cite{Campbell:2011bn}}   & $WW +$ jets  & $124.31$& NLO\\ 
                  & $WZ +$ jets  & $51.82$ & NLO\\ 
                   & $ZZ +$ jets  &  $17.72$ & NLO\\ \hline
\multirow{3}{*}{Single $t$~\cite{Kidonakis:2015nna}}  & $tW$  &  $83.10$ & N$^2$LO \\ 
                   & $tb$  & $248.00$ & N$^2$LO\\ 
                   & $tj$  &  $12.35$ & N$^2$LO\\  \hline
$tt$~\cite{Muselli:2015kba}  & $tt +$ jets  & $988.57$ & N$^3$LO\\ \hline
\multirow{2}{*}{$ttV$~\cite{Kulesza:2018tqz}} & $ttZ$  &  $1.05$ &NLO+NNLL \\ 
                   & $ttW$  & $0.65$& NLO+NNLL \\ \hline
\end{tabular*}}
\caption{The significant SM background processes relevant to our signal topology (adapted from \cite{Bhaskar:2020gkk}). We used the highest-order QCD cross sections available in the literature for the background processes.}
\label{table:Backgrounds}
\end{table}
\section{LHC phenomenology}\label{sec:pheno}
\noindent
To study the prospects of our model at the LHC, we consider the $b\bar b\tau^+\tau^-$ channel where one of the Higgs boson decays to a $b\bar b$ pair and the other to a pair of $\tau$'s. We use {\sc MadGraph5}~\cite{Alwall:2014hca} to generate the signal and background events at the tree level. We implement the following effective Lagrangian  in {\sc FeynRules}~\cite{Alloul:2013bka} to generate the {\sc Universal FeynRules Output}~\cite{Degrande:2011ua} model files:\footnote{Since we consider the $\phi$ to be about as heavy as the LQ running in the loops, one must be careful in using the effective Lagrangian for computing cross sections. Therefore, we use it only to simulate the kinematic distributions of the Higgs pair, not the signal cross sections.}
\begin{align}\label{eq:LagPhiInt}
	\mathcal{L}_{int}\ \supset&\ \mu_{h\phi}\left(H^\dagger H\right)\phi + \frac{1}{\Lambda_g} G_{\mu\nu}^a G^{\mu\nu a}\phi 
	+ Y_{d\bar d\phi} \bar{d}_L d_R \phi\nonumber\\
	&\ + Y_{s\bar s\phi} \bar{s}_L s_R \phi + Y_{b\bar b\phi} \bar{b}_L b_R \phi + {\rm H.c}.
\end{align}
where, $\mu_{h\phi}$ and $\Lambda_{g}$ are the dimension-one couplings of the singlet $\phi$ to the SM Higgs and gluons, respectively. To estimate the signal strength for our LHC analysis, we use the K-factors, K$_{qq~(gg)}= 1.3~(1.72)$ from Ref.~\cite{Alasfar:2019pmn}. We incorporate the higher order corrections to the SM background processes as well. We multiply the LO cross sections with the appropriate QCD $K$-factors known in the literature (see Table~\ref{table:Backgrounds}). We use the NNPDF2.3LO~\cite{Ball:2012cx} parton distribution functions with the default renormalisation and factorisation scales. For showering and hadronisation, these parton level events are passed through {\sc Pythia6}~\cite{Sjostrand:2006za}. The detector simulations are performed in {\sc Delphes 3.5.0}~\cite{deFavereau:2013fsa} using the \textsc{Atlas} card, incorporating the latest $b$ quark and $\tau$ lepton identification efficiencies\footnote{We take the $b$-quark tagging efficiency to be a constant beyond the $p_T$ regime mentioned in Ref.~\cite{ATLAS:2019bwq}.}~\cite{ATLAS:2019bwq,ATLAS:2015xbi}. We use the FastJet package~\cite{Cacciari:2011ma} for jet clustering with the anti-$k_T$ algorithm~\cite{Cacciari:2008gp} with the radius parameter $R = 0.4$. 

A simple kinematic-cuts-based analysis is not very promising in this case, as the SM backgrounds are quite significant. Hence, we employ a multivariate analysis that uses Boosted Decision Trees (BDTs) to estimate the signal significance at the HL-LHC.
\medskip

\begin{figure}
\captionsetup[subfigure]{labelformat=empty}
\includegraphics[width=0.48\textwidth]{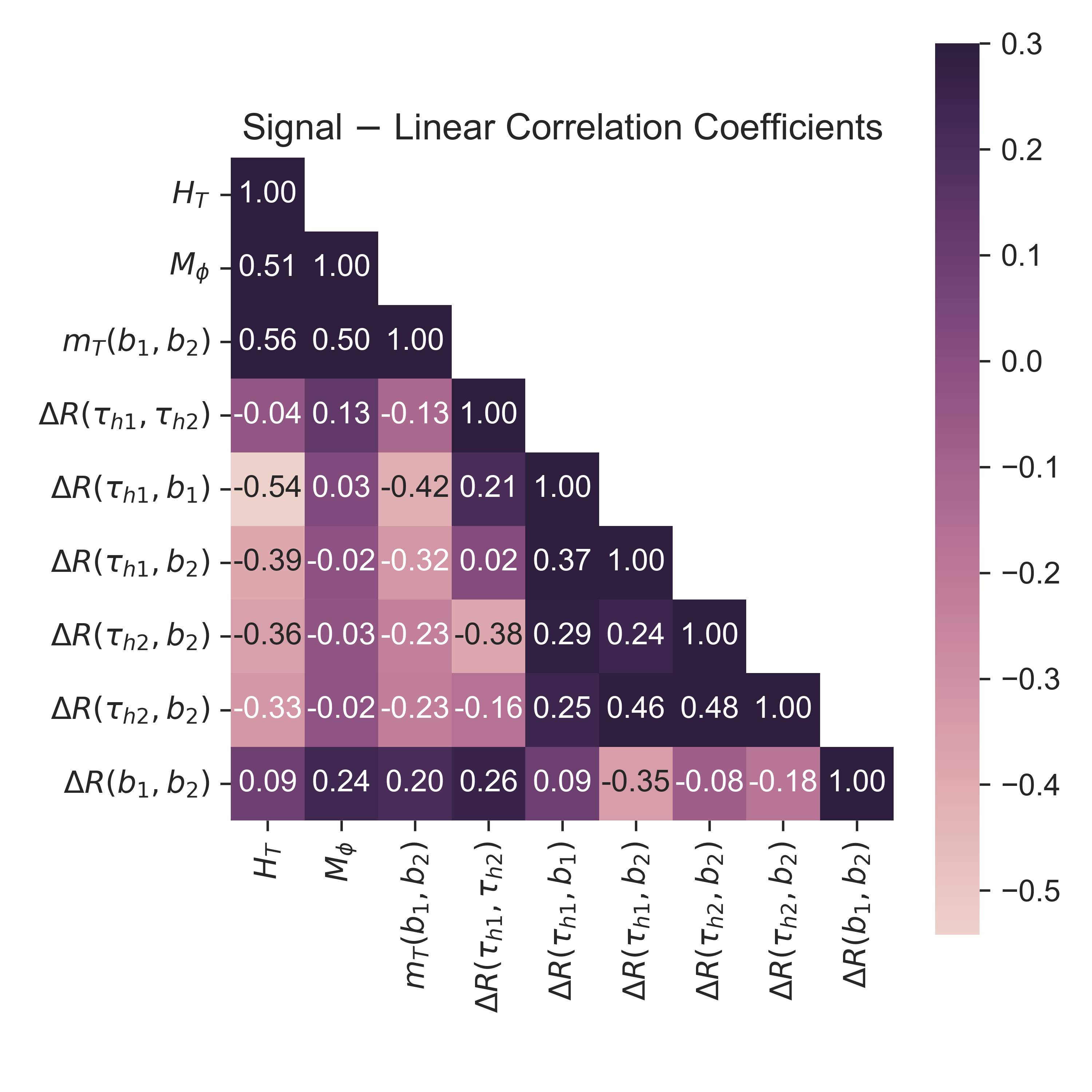}
\caption{Pearson’s linear correlation coefficients in the signal for the inputs chosen for BDT analysis.}
\label{fig:BDT_varCorr}
\end{figure}
\begin{table}[!t]
\begin{centering}
\begin{tabular*}{\columnwidth}{l @{\extracolsep{\fill}} l}
\hline 
Parameter & Optimised choice\tabularnewline
\hline 
\hline 
NTrees & $500$\tabularnewline
MinNodeSize & $5.0$\%\tabularnewline
MaxDepth & $5$\tabularnewline
BoostType & AdaBoost\tabularnewline
AdaBoostBeta & $0.5$\tabularnewline
UseBaggedBoost & True\tabularnewline
BaggedSampleFraction & $0.5$\tabularnewline
SeparationType & GiniIndex\tabularnewline
nCuts & 20\tabularnewline
\hline 
\end{tabular*}
\par\end{centering}
\caption{Summary of optimised BDT hyperparameters.}\label{table:bdt_params}
\end{table}

\begin{table}[t!]
\centering{\linespread{2}
\begin{tabular*}{\columnwidth}{l @{\extracolsep{\fill}} rlr }
\hline
\multirow{2}{*}{Feature} & \multirowcell{2}{Method-unspecific \\ Separation} &  \multirow{2}{*}{Feature} & \multirowcell{2}{ Method-specific \\ Ranking} \\
&&&\\\hline\hline
$\Delta R(b_1, b_2)$ & $0.4968$ &  $ \Delta R(b_1, b_2)$ &  $0.1969$ \\
$\hat M_{\phi}$ & $0.4209$ &  $H_T$ & $0.1601$ \\
$\Delta R(\tau_{h1}, \tau_{h2})$ & $0.3908$  & $\hat M_{\phi}$ & $0.1302$ \\
$\Delta R(\tau_{h1}, b_2)$ & $0.3864$ &  $m_T(b_1, b_2)$ & $0.1120$ \\
$m_T(b_1, b_2)$ & $0.3386$ &   $\Delta R(\tau_{h1}, \tau_{h2})$ & $0.1015$ \\
$\Delta R(\tau_{h2}, b_1)$ & $0.3141$ &  $\Delta R(\tau_{h1}, b_2)$ & $0.0913$ \\
$\Delta R(\tau_{h2}, b_2)$ & $0.1938$ &  $\Delta R(\tau_{h1}, b_1)$ & $0.0902$ \\
$\Delta R(\tau_{h1}, b_1)$ & $0.1287$ &  $\Delta R(\tau_{h2}, b_1)$ & $0.0734$ \\
 $H_T$ & $0.0578$ & $\Delta R(\tau_{h2}, b_2)$ & $0.0443$ \\
\hline 
\end{tabular*}}
\caption{Summary of method-unspecific and method-specific ranking for the input features at the benchmark mass point.}
\label{table:feature_sepn}
\end{table}
\begin{figure*}
\captionsetup[subfigure]{labelformat=empty}
\subfloat[\quad\quad(a)]{\includegraphics[width=0.49\linewidth]{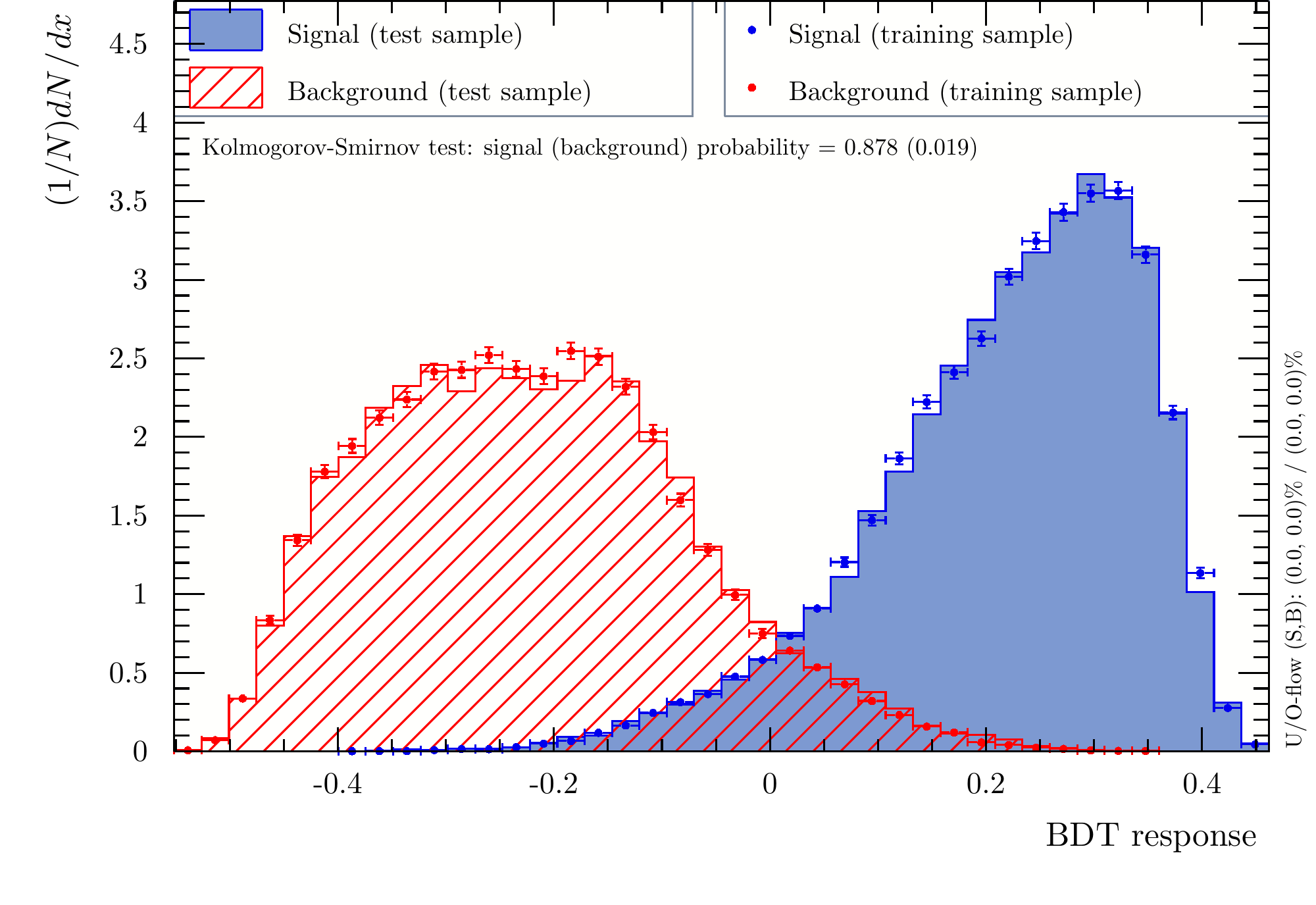}\label{fig:1TeV_ClassCut}}\hspace{0.2cm}
\subfloat[\quad\quad(b)]{\includegraphics[width=0.49\linewidth]{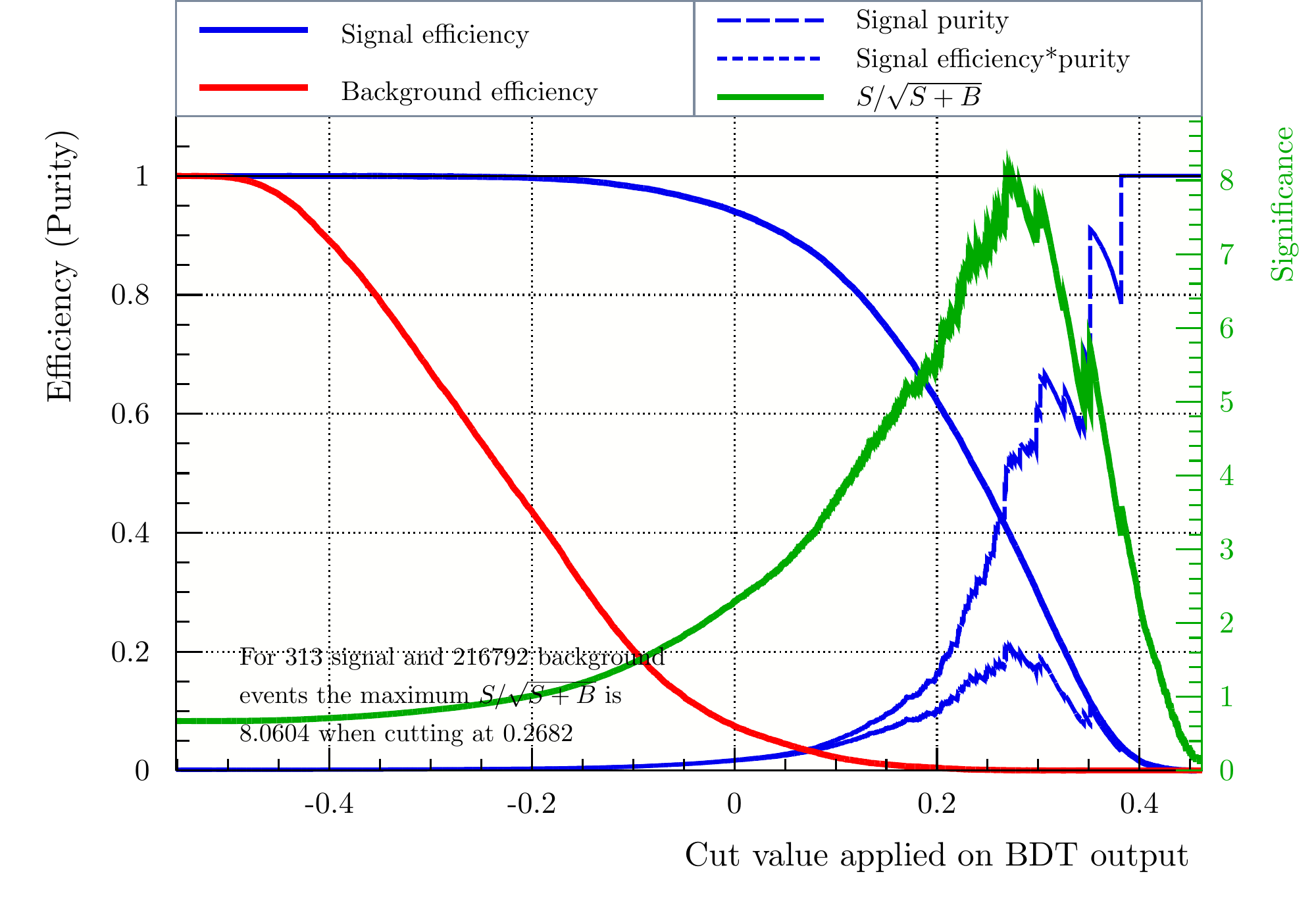}\label{fig:KSTest}}
\caption{BDT Classifier information for $M_\phi = 1$ TeV: (a) normalised BDT response for the signal and background distributions (training and test samples), (b) classifier cut efficiency as a function of the BDT-cut values.}
\label{fig:bdt_info}
\end{figure*}
\begin{figure*}
\captionsetup[subfigure]{labelformat=empty}
\subfloat[\quad\quad(a)]{\includegraphics[width=0.45\linewidth]{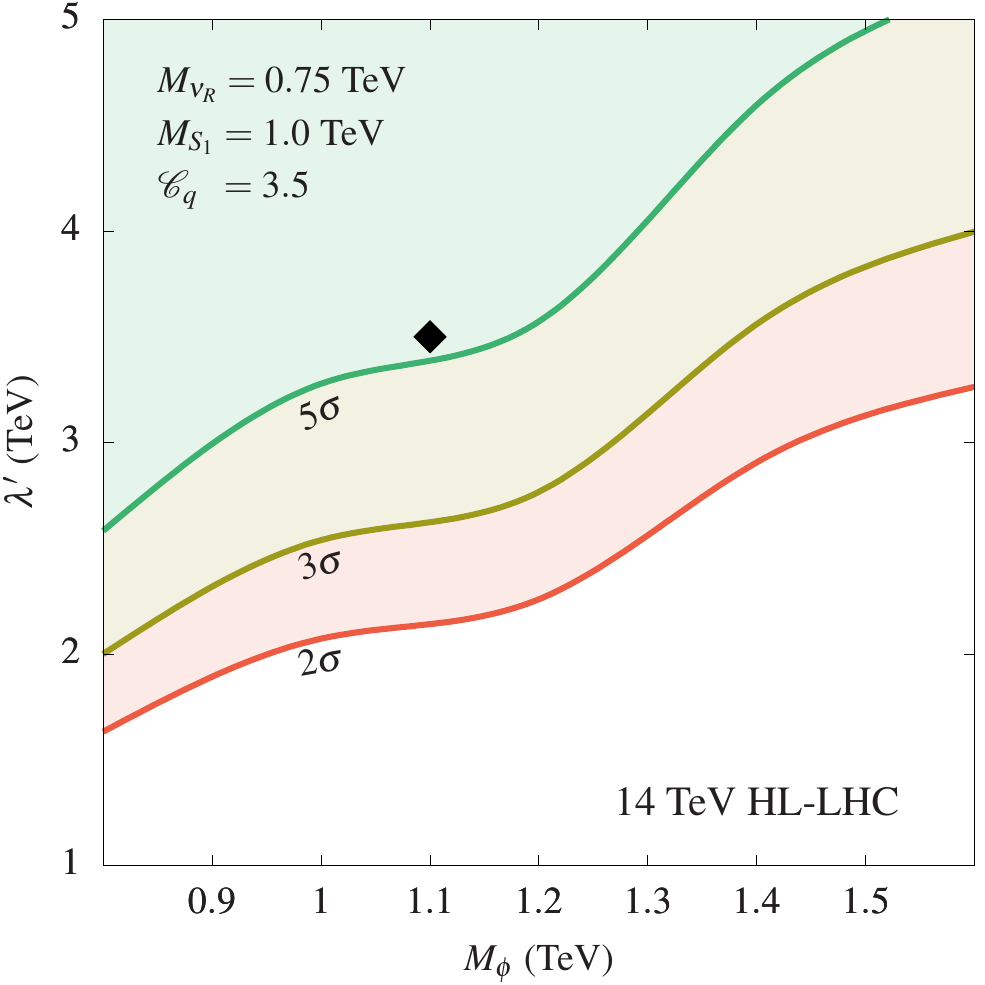}\label{fig:lambda_Mphi}}\hspace{1cm}
\subfloat[\quad\quad(b)]{\includegraphics[width=0.45\linewidth]{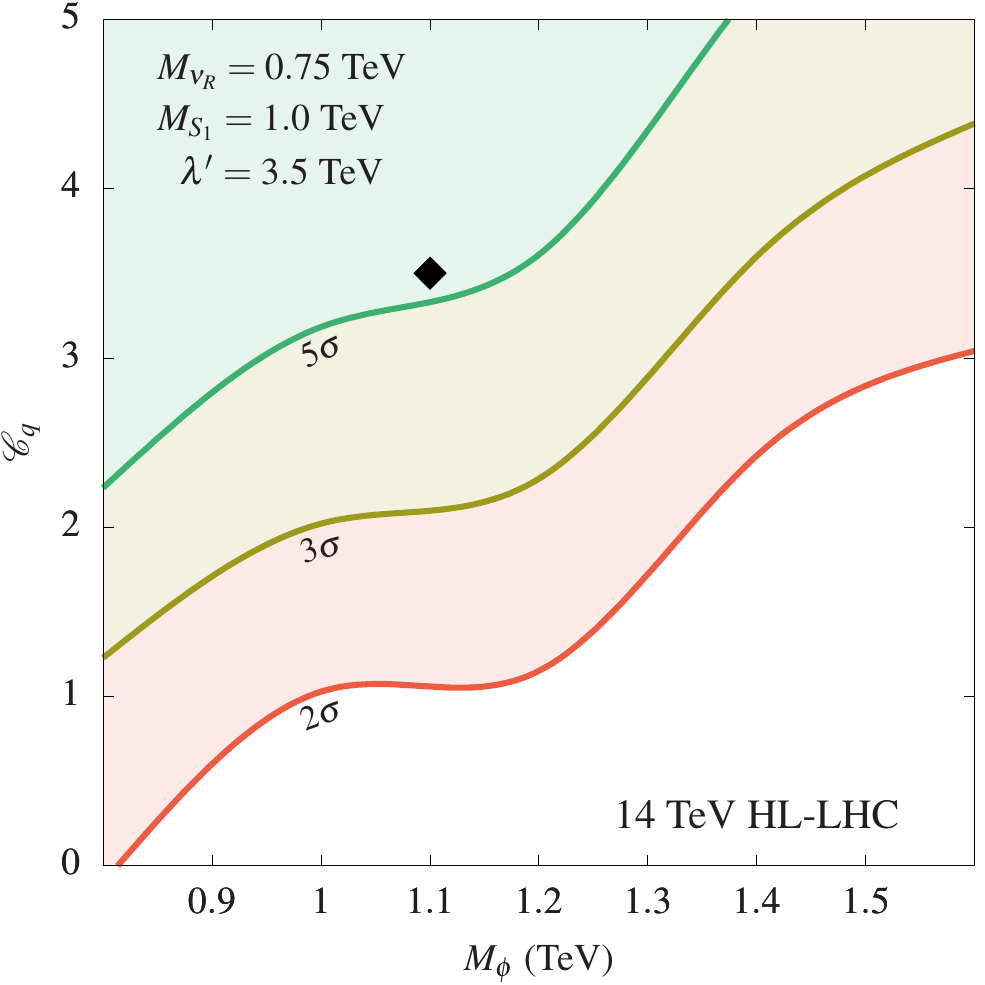}\label{fig:gR_Mphi}}
\caption{HL-LHC ($3$ ab$^{-1}$) projection plots: $5\sigma$ (discovery), $3\sigma$ and $2\sigma$ contours in the (a) $M_\phi-\lambda'$ and (b) $M_\phi-\mathcal{C}_q$ planes. Regions above the $2\sigma$ ($5\sigma$) lines can be excluded (discovered) at the HL-LHC. The black diamond represents the  parameter point marked in Fig.~\ref{fig:CSMphi}: at this point, $\sigma(pp\to \phi)$ in our model fits the slight excess seen by ATLAS~\cite{ATLAS:2021tyg}. At $14$ TeV, it corresponds to a $\phi$ production cross section of about $16.7$ fb.}
\label{fig:lambdaMass}
\label{fig:sig_eff}
\end{figure*}

\noindent{\bf A multivariate analysis:}
We generate the process $p p \to \phi \to h h \to b\bar b\tau^+\tau^-$ and select the events with the following event-level criteria:\
\begin{itemize}
  \item Cut-$1$: At least $2$ jets tagged as $\tau_h$ leptons (hadronic $\tau$'s)
  \item Cut-$2$: At least $2$ jets tagged as $b$ quarks
  \item Cut-$3$: $\Delta\phi(\tau_{h1}, b_1) > 2$, i.e. a high $\phi$-separation between the leading-$p_T$ $\tau$ lepton and $b$ quark.
\end{itemize}
\noindent These cuts lead to a significant drop in the background events and a moderate drop in the signal events. After these cuts, the efficiencies of quark mediated and gluon mediated signal events are similar ($\approx 4.5\%$). Demanding  two $b$ quarks essentially eliminates the $W$ backgrounds. We find that only the $t\bar{t}$ backgrounds remain significant after these cuts. Using the event-level information and the kinematic features of identified objects, we define the following basic set of quantities: 
\begin{align}
  & H_T, \hat{M}_{\phi},\, m_{T}(b_1, b_2),  \Delta R(\tau_{h1}, \tau_{h2}), \, \Delta R(b_1, b_2), \nonumber\\
  &\Delta R(\tau_{h1}, b_1),  \Delta R(\tau_{h1}, b_2), \,  \Delta R(\tau_{h2}, b_1), \,   \Delta R(\tau_{h2}, b_2) 
  \label{eq:bdtVar}
\end{align}
where $m_T(b_1, b_2)$ is the transverse mass of the $bb$-system (both transverse and invariant mass of the $bb$-system have similar distributions and are highly correlated, we pick $m_T$ since it has slightly better separation visibly), $\Delta R(x,y)$ is the separation between $x$ and $y$ in the $\eta-\phi$ plane, $\hat{M}_{\phi}$ is the estimator of the invariant mass of the singlet $\phi$ constructed from the tagged $\tau_h$-leptons and $b$-quarks and $H_T$ is the sum of the $p_T$ of all jets. We picked these features from a larger set of kinematic variables created from the identified objects after dropping the highly correlated ones (linear correlation $>70\%$ in the signal)---we pick the one with the greatest separation out of a highly co-related pair. The Pearson correlation coefficients between the final set of features are shown in Fig.~\ref{fig:BDT_varCorr}. 

We use these features as inputs for the BDT analysis with the adaptive boosting (AdaBoost) optimisation. After obtaining statistically independent event samples for all the signals and background processes considered, we take $30\%$ of the dataset for testing and the rest for training the algorithm. Since there are multiple processes in each of the signal and background classes, we weigh each by the ratio of the expected number of events from the process at the HL-LHC to the number of events fed into the BDT algorithm. We choose the benchmark mass, $M_{\phi} = 1$ TeV to optimise the BDT algorithm, which we implement using the TMVA~\cite{Hocker:2007ht} package in ROOT. The Kolmogorov-Smirnov test ensures that the  signal and background distributions are well separated---see Fig.~\ref{fig:bdt_info} for the normalised BDT response and classifier efficiency. The set of optimised BDT hyperparameters are given in Table~\ref{table:bdt_params}. To show the discriminating power of this set of features in the context of BDTs, we list two measures for each feature: method-unspecific separation and method-specific ranking. The method-unspecific separation which is defined as,
\begin{equation}
    \left<S^{2}\right>=\frac{1}{2}\int dy \frac{\left(\hat{y}_{S}\left(y\right)-\hat{y}_{B}\left(y\right)\right)^{2}}{\hat{y}_{S}\left(y\right)+\hat{y}_{B}\left(y\right)}\label{eqn:tmva_separation}
\end{equation}
where $\hat{y}_{S}$ and $\hat{y}_{B}$ are the probability density functions of the signal and background respectively for a particular feature $y$. For signal and background distributions with no overlap, it is unity---features with higher method-unspecific separation values are better at resolving the signal events from the background. Method-specific ranking is derived by counting how often the variables are used to split decision tree nodes, and by weighting each split occurrence by the separation gain-squared it has achieved and by the number of events in the node. We show both these quantities in decreasing order of discriminating power of the features in Table~\ref{table:feature_sepn}. The significance is computed using the formula, $Z = \mathcal{N}_s/\sqrt{\mathcal{N}_s + \mathcal{N}_b}$ where $\mathcal{N}_s$ ($\mathcal{N}_b$) is the number of signal (background) events surviving at the HL-LHC after the BDT cut. We use the optimised analysis setup to study the collider reach for different masses of $\phi$ in the range $0.8-1.6$ TeV, for the benchmark $\lambda'$ and $\mathcal{C}_q$. Based on our parametrisation, the number of signal events scales with $\lambda'$ and $\mathcal{C}_q$ as,
\begin{equation}
    \mathcal N_s = \left(\frac{\lambda'}{\rm TeV}\right)^2 \left( \bar{\mathcal N}_s^{gg} + \,(\mathcal{C}_q)^2\bar{\mathcal N}_s^{qq}\right), \label{eq:N_Sig}
\end{equation}
where $\bar{\mathcal N}_s^{gg(qq)}$ is the number of $gg(qq)$-initiated events surviving the BDT cut at $3$ ab$^{-1}$ when $\lambda^\prime=1$ TeV (and $\mathcal C_q=1$). We present the $5\sigma$ and $2\sigma$ limits in terms of $\lambda'$ and $\mathcal{C}_q$ in Fig.~\ref{fig:lambdaMass}. One can easily translate these limits in terms of the signal cross section, correcting for the efficiency of each process. We see that for $\mathcal C_q=3.5$, $\lambda^\prime=3.5$ TeV, a $1.1$ TeV $\phi$ (i.e., the parameters to accommodate the current ATLAS excess) can be discovered at the HL-LHC---the signal significance for this point is slightly above $5\sigma$.

\section{Summary and conclusions}\label{sec:conclu}
\noindent
In this letter, we have studied the resonant di-Higgs production mediated by a SM-singlet scalar, $\phi$ at the LHC. In our model, the only SM particle the singlet directly couples to is the Higgs. It couples to $S_1$, the charge $-1/3$ weak-singlet scalar Leptoquark, through which it gains an effective coupling with gluon pairs. It also couples to down-type quark-antiquark pairs through loops involving $S_1$ and $\nu_R$. We showed in an earlier paper~\cite{Bhaskar:2020kdr} that the quark fusion contribution to $\sigma(pp\to\phi)$ could be comparable, or even more significant, than the gluon fusion channel for a suitable choice of parameters. Moreover, since the singlet $\phi$ has tree-level couplings to the Higgs, it dominantly decays to a pair of Higgses. As a result, we have studied an interesting setup where the $\phi$ can be substantially produced at the LHC through both quark and gluon fusion channels and decays to the di-Higgs final state. The parameter ranges for our study were motivated from the  ($3.2\sigma$ local significance) excess in the di-Higgs spectrum around $M_{hh}\approx 1.1$ TeV observed by the ATLAS collaboration~\cite{ATLAS:2021fet,ATLAS:2021tyg} at the $13$ TeV LHC.

We have investigated the prospects of resonant di-Higgs production in our model at the HL-LHC. For this purpose, we have chosen the $b\bar b\tau^+\tau^-$ final state (in which the ATLAS excess was observed~\cite{ATLAS:2021fet}). To overcome the huge SM background in this channel, we performed a multivariate analysis using Boosted Decision Trees to estimate the signal significance at the HL-LHC. Our study has shown that if we choose the model parameters to account for the excess at the $13$ TeV LHC, the HL-LHC can discover such a resonance. However, there are other variants of multivariate analysis (e.g., BDTs with different boosting techniques, deep-neural-network-based analyses, etc., for e.g. see \cite{Adhikary:2020fqf,Adhikary:2018ise}) that could lead to a better signal yield at the HL-LHC. Moreover, with better $b$- and $\tau$-tagging efficiencies in the future, one could improve this further. A similar study can also be performed in the $bbbb$ mode, which we postpone for a future study. We have presented the results of our study for a wide range of model parameters that can be easily translated to cross sections and, hence, useful for other similar models.

\relax
\section{Acknowledgments}
\noindent 
Our computations were supported in part by SAMKHYA: the High Performance Computing Facility provided by the Institute of Physics (IoP), Bhubaneswar, India. C. N. is supported by the DST-Inspire Fellowship. 
\vfill
\bibliographystyle{JHEPCust.bst}
\bibliography{DiHiggs_LQ}

\end{document}